# Gravity-Superconductors Interactions as a Possible Means to Exchange Momentum with the Vacuum[1]

**Giovanni Modanese**

**Free University of Bolzano-Bozen**
**Faculty of Science and Technology**
**P.za Università, 5 – 39100 Bolzano, Italy**

E-mail: Giovanni.Modanese@unibz.it

**ABSTRACT:** We report on work in progress in quantum field theory about possible interactions between coherent matter, i.e. matter described by a macroscopic wave function or a classical field, and a certain class of vacuum fluctuations, called "zero-modes of the Einstein action". These are little-known virtual masses present in the vacuum state of quantum gravity. A couple of equal masses of this kind can be excited by an oscillating coherent source with frequency $f$ and decays to its ground state emitting a virtual graviton, which can propagate and transfer momentum $p$ to ordinary matter. The virtual masses recoil in the emission, and this amounts to a transfer of momentum $-p$ to the vacuum; this momentum can be passed in turn to some matter, or not. The energy $hf$ for the process does not come from the vacuum, but from the coherent source. The ratio $hf/p$ is of the order of 1 m/s. This model was developed to explain experimental results showing the emission of anomalous high-momentum radiation from certain superconductors, sometimes with a strong recoil of the emitters. The recoil is energetically quite efficient, at least at small power, and could be exploited for propulsion. It has not been tested in space, however, and our model cannot yet predict if the recoil is affected by the presence of near matter. (Another model predicts that it is not.) We also briefly mention a possible application of the anomalous radiation itself and we evaluate the (large) electric and magnetic field strength needed to produce an effect equivalent to that of a superconducting emitter.

## 1. INTRODUCTION

One of the main limitations to the efficiency of propulsion methods in space is the need for a reaction mass: in addition to a source of energy, spacecraft must carry on board a sufficient amount of "passive" material, to be expelled in order to obtain a reaction force. This is true also for small boosters employed, for instance, in the adjustment maneuvers of satellites over a lifetime of years: they become useless after exhausting their propellant, even though the main system which supplies energy to the satellite could still be able to energize them. Several authors have speculated on the possibility to obtain "propellantless propulsion" by exploiting some exotic properties of spacetime (warp drive, vacuum energy, etc.). The subject is now well described also at divulgation level, in books like [1]. These books are based, of course, on established notions of General Relativity and quantum physics, and offer the sobering conclusion that although certain exotic processes are possible in principle, their concrete application to propulsion is not realistic [2].

---

[1] Submitted to J. of Space Exploration, Special Issue "Spaceflight Perspectives from Novel Concepts of Spacetime, Gravitation and Symmetries", Ed. J.H. Hauser.

There exist, however new theories and new experimental facts being reported, and we think it is important to analyze and discuss them. This is not yet "official mainstream science", blessed by the mayor incumbent academicians, because the data are still scarce. Nevertheless, it is taken seriously by those who made these discoveries and are testing and publishing them (with some difficulties). The new findings may also happen to be interesting for visionary entrepreneurs. We hope, however, that any novel physical principles will not be confined in patents or proprietary research, but gradually brought into the mainstream, clarified and checked by more researchers.

I would like to state at the very beginning what I think is possible and what I think is not, according to my personal feeling; in the paper I will give arguments for this, as far as allowed by the limited data available. I think that processes are possible, in which coherent matter exchanges energy and momentum with a certain class of vacuum fluctuations (not the familiar vacuum fluctuations of Quantum Electrodynamics or their analogues in the Standard Model, but anomalous and little known vacuum fluctuations of Quantum Gravity [3]). This may allow some form of propulsion without reaction mass. I believe, however, that the energy necessary for the propulsion cannot be extracted from the vacuum fluctuations, but must come from a conventional on-board source. In very simple terms, I think the situation reminds the working principle of a jet engine, which propels itself in one direction by boosting air in the opposite direction: the energy is provided by the fuel, while the momentum is balanced by the air molecules passing through the engine and eventually "dispersed" in the atmosphere.

The experimental evidence to which I make reference (the discharge experiments by E. Podkletnov and C. Poher [4,5,6]) gives different results for momentum exchange, depending on the conditions. In both the devices of Podkletnov and Poher, the momentum imbalance due to the exchange with the vacuum can occur in principle in two parts of the system: in the targets hit by the anomalous radiation and in the recoil of the emitter (Fig. 1). The observations show that Podkletnov measures a large momentum in the targets, but no recoil of the emitter; Poher has comparatively little momentum in the targets, and large recoil in the emitter.

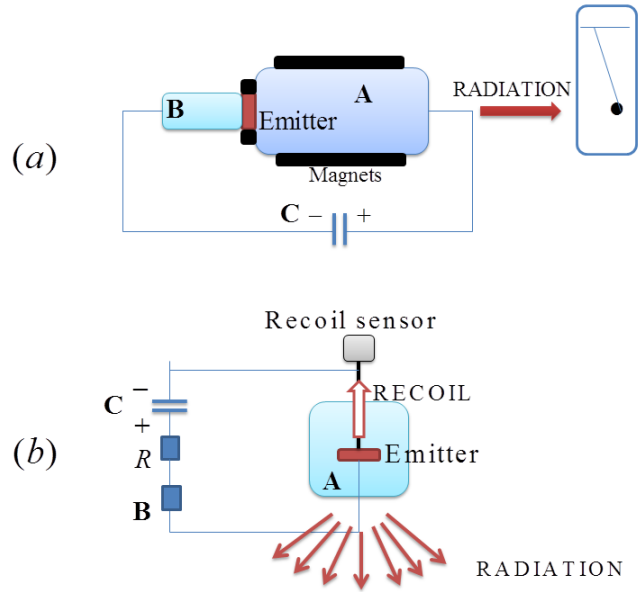

**Figure 1** – **(a)** Scheme of the device by E. Podkletnov for high-voltage discharges through a superconducting YBCO emitter. The emitter (diameter 10 cm) and the vacuum chamber (A, diameter ca. 1 m) are surrounded by electromagnets. The emitter is cooled by lateral contact with a liquid helium reservoir (B). A Marx generator (C) produces an over-damped high voltage pulse of at least 500 kV. The circuit has a distributed inductance of ca. 10 to 15 µH, but no load resistance. The emitted anomalous radiation propagates to a large distance in a collimated beam. Its effects are measured by the impact on small ballistic pendulums of variable mass and composition.
**(b)** Scheme of the device by C. Poher for medium-voltage discharges in a superconducting emitter. The YBCO emitter (diameter 1 - 8 cm) is suspended in a liquid nitrogen bath (A) and mechanically coupled to a recoil sensor. The discharge is produced by a capacitors bank (C) with a max. voltage of 4.5 kV and switched by a thyristor (B). There is a distributed inductance in the circuit of the order of 1 µH and a load resistance of the order of 0.1 Ω. The emitted radiation is measured in a shielded box of sensors placed approx. 25 - 50 cm below the emitter and has an angular distribution which depends on the emitter type. A strong recoil is always detected in the opposite direction to the radiation.

These observations suffer from several limitations. (1) The targets have small mass: a few grams for Poher accelerometers (actuators P888-91 from PI), up to 50 g for Podkletnov ballistic pendulums (although he informally reported that heavier targets, up to approx. 1 kg, were displaced by force beams generated with very large voltage). Note that for such small targets, the acceleration appears to be independent from the mass. (2) Podkletnov setup does not comprise any device which could display and amplify a recoil. The recoil might be present, but concealed by the large mass of the emitter and of the discharge chamber rigidly connected to it; or there might be a recoil force internal to the setup, causing only a strain between components which are rigidly connected, like for instance if the emitter "pushes" on the discharge chamber. (3) Poher's force beam is diverging, and his detectors only cover a small angle at a time; there are no large-angle integrating detectors.

In our microscopic model (to be summarized in Sect. 2), the anomalous radiation generated in the superconducting emitters of Podkletnov and Poher originates from the decay of strong gravitational fluctuations which are excited by the interaction with the macroscopic wavefunction of the emitter [7,8]. The anomalous radiation propagates towards mobile targets which absorb it and acquire its momentum, while fixed massive targets or screens do not absorb it. The emitting vacuum fluctuations in the superconductor recoil, but due to their very nature of "Lorentz-invariant objects" they cannot pass their recoil momentum to other vacuum fluctuations, which always appear to them as uniform and isotropic; under this respect, they behave very differently from air molecules accelerated by a jet engine, which pass their momentum to other molecules by scattering. The vacuum fluctuations can in principle pass their momentum to ordinary matter, but the scattering cross section of this process is exceedingly small (Sect. 4). It is possible that the cross section of their scattering with coherent matter is larger. In fact, we have proven in [7] that the excitation probability of the vacuum fluctuations by interaction with ordinary matter is very small, while the same probability becomes relevant when a coherent wave function is involved; the same could happen for the scattering cross section, but there are no simple arguments in favor or against this hypothesis. In any case, the detailed phenomenology of the process would be complicated, depending on whether the anomalous radiation is generated "in front" of the superconductor or behind it (Sect. 3).

In Sect. 2 we continue the discussion of the momentum balance in the two experiments and we summarize the main ideas of our theoretical model. In Sect. 3 we analyze in particular the theoretical predictions concerning the recoil of the emitter, showing that there are several different possible alternatives. In Sect. 4 we present further details and improvements of the theoretical model in general. Sect. 5 comprises our Conclusions and some updates.

## 2. ANOMALOUS GRAVITY-SUPERCONDUCTORS INTERACTIONS

In the last years the subject of "Gravity-Superconductors interactions" has attracted considerable interest (compare the ebook [9], with an historical introduction and extensive references in the first chapter [10], entirely accessible on Google Books). A number of recent experiments show an apparent generation, in certain laboratory conditions and from certain superconductors, of gravitational-like fields which are clearly outside the predictions of General Relativity. In the experiments by Tajmar et al. [11,12] the qualitative features of the field are familiar: it looks like a gravitomagnetic field, but several orders of magnitude stronger than predicted by the usual Einstein-Maxwell equations. A possible theoretical explanation was offered by Hauser and Dröscher, based on the concept of electro-gravitational conversion at low temperatures within the extended Heim field theory [13]. This theory requires an extension of the fundamental symmetries of particles physics. Possible applications to propulsion have been discussed in [14]. Our theoretical model and that of Poher and Marquet have not been used to explain Tajmar experiments.

In the high-voltage experiments by E. Podkletnov [4] the field generated by the superconductor is very different from any known classical field and, besides having an unusual strength, does not satisfy any field equation compatible with General Relativity. It looks like a focused beam of radiation with very large $p/E$ ratio ($p/E \approx 1$ s/m). It may be called "gravitational" because it exerts on the targets a force proportional to their mass. Possible applications of this beam to beamed propulsion were suggested already in [15], but the force/mass proportionality might be valid only in a limited range and the energetic efficiency appears to be low. More realistic conceivable applications comprise the utilization of the mechanical effects of the beam (coupled to an array of piezoelectric sensors) for scanning materials or biological tissues (Fig. 2).

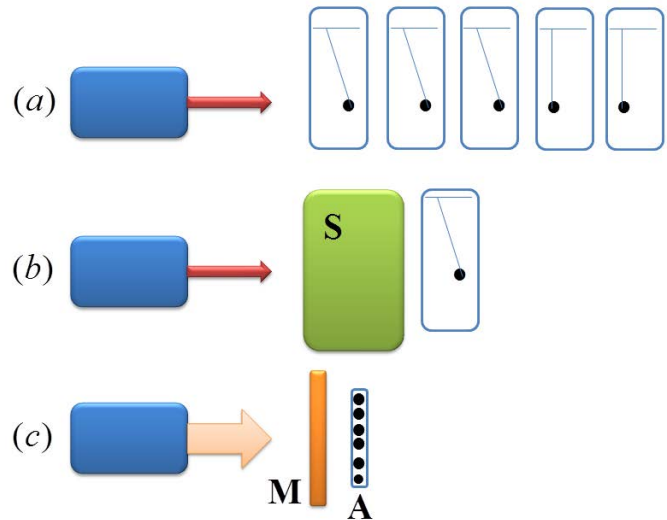

**Figure 2** – "Total absorption hypothesis" of Podkletnov beam and possible application for materials scansions. **(a)** Total absorption hypothesis: the maximum energy of the emitter and the target velocity are fixed, therefore there is a maximum target mass which can be accelerated by the beam, absorbing all its energy. (Note that this hypothesis is consistent with our theoretical model, but not with the Universons model of Poher and Marquet.) **(b)** For this reason the beam can penetrate large massive and rigid screens (S): the total energy available would not be enough to accelerate them, therefore any energy transfer to them is inhibited. **(c)** A low-power beam with large cross-section has low penetration ability and could be used as an alternative to ultrasound scans for inspection of a material (M). Detection requires a 2D array of piezoelectric sensors (A). The short wavelength of the virtual gravitons (1 micrometer or less) would allow a high resolution.

Podkletnov's high-voltage experiment, however, is difficult to replicate [16]. Since the partial replications of a previous experiment by Podkletnov gave negative results [10], some skepticism has arisen about these results. Recent work by C. Poher with a device which generates discharges in an emitter at lower voltage, lends more credibility to Podkletnov and opens novel prospects. A detailed comparison of the two experiments was given in [6]. The generation conditions for Poher are somewhat different: beside the lower applied voltage (4 kV vs. 500 kV), remarkable are the longer duration of the pulses and the absence of a discharge chamber. The microstructure of the emitter is also different in the two cases. While Podkletnov's radiation beam is collimated, Poher's beam is more or less diverging, depending on the emitter type. In coincidence with the radiation emission Poher measures a strong recoil of the emitter, with maximum momentum of the order of 1 kg·m/s. Typical energy efficiency values for powerful multi-layer emitters are 10 g·(m/s) per $cm^2$ of emitter surface and per J of electric current.

## 2.1 Momentum and energy balance

The striking similarities and differences between these experiments and their results represent a challenge for any comprehensive analysis. This applies in particular to the explanation of the recoil. Furthermore, different theoretical models make different predictions on the possible use of the recoil phenomenon for propulsion.

For instance, an explanation of the anomalous radiation and of the recoil must include a balance of the total momentum. A first natural assumption is that the momentum carried by the radiation is balanced by the recoil momentum of the emitter. But can a "momentum carried by the radiation" really be defined, independently from the targets? Or should one only speak of a momentum transferred to the targets, and therefore depending on the available targets? The radiation beam appears to have an energy/momentum ratio incompatible with the hypothesis that it is made of real particles. It seems therefore that it only makes sense to speak of momentum transferred by virtual particles and it is impossible to consider the radiation independently from the target. Furthermore, since the momentum imparted by the radiation on a target is proportional to the mass of the target, it is not obvious that one can detect the same momentum per unit surface, no matter how many detectors one places, and no matter how much they weigh. All this might imply that the recoil of the emitter depends to some extent on the target. There would also exist a maximum target mass, such that for larger masses the target acceleration would not be constant, but would decrease and tend to zero (compare Fig. 2 and [17]).

Also the analysis of the energetic balance depends on the theoretical model adopted. As mentioned in the Introduction, I believe that the process of emission, propagation and absorption of the anomalous radiation involves an exchange of momentum with the vacuum, but not an extraction of energy from the vacuum. According to the Universons model of Poher [18], on the contrary, there should be an excess energy. Poher has hypothesized that the recoil energy of the emitter might exceed in certain conditions the electric energy $E_{el}$ supplied by the external circuit, and would therefore be extracted from the universons background. This hypothesis would be supported by the fact that the measured recoil energy of the emitter is proportional to the square of $E_{el}$. By extrapolation, one would obtain conditions of over-unity energy balance. Real measurements in these conditions are not yet been reported, however, so the quadratic dependence might actually fail at some point. A similar over-unity conjecture concerns the mass of the emitter, since the recoil energy is inversely proportional to this mass. It is not certain, however, whether one could make the emitter lighter without affecting its emission rate, for instance by using lighter materials for the non-superconducting parts.

It should also be mentioned that the Universons model by Poher gives a different picture of the whole phenomenon. This model postulates the existence of a powerful background flux of real particles, which can be "intercepted" by the superconducting emitter. The emitter is able to divert a small part of the flux and extract some energy from it. This interpretation allows to circumvent a considerable conceptual difficulty, namely to explain how the observed anomalous radiation can convey to the targets a momentum which is much larger than the radiation pressure momentum $p=E/c$. In the Universons model, the "transmission balance" is not limited to emitter and radiation, but also encompasses the surrounding background. Roughly speaking, a kind of vacuum pressure is involved, which is able to transfer much more momentum than could be carried by single particles. In this picture, the Universons of the anomalous beam are not just absorbed in the target, but absorbed and quickly re-emitted. Poher's model is essentially "classical and global", while our model is "quantum-mechanical and local".

## 2.2 Our theoretical model in short

Our theoretical model is based on vacuum fluctuations of two kinds: virtual gravitons (which make up the anomalous radiation beam) and massive gravitational "zero-modes", an entirely novel kind of fluctuations which are supposed to emit the virtual gravitons. Each elementary process of absorption of a radiation quantum in a target corresponds to an elementary emission process in the decay of a zero-mode. The zero-modes are excited by the interaction of the vacuum with the wave function of the superconductor. The whole process is quantum-mechanical, and the radiation is only virtual. The energy is supplied by the superconductor and therefore by the electric circuit which generates the supercurrent. The momentum acquired by the target causes a recoil of the zero-mode, and is therefore transferred to the vacuum, unless the zero-mode is scattered by coherent matter (see discussion in Sect. 3).

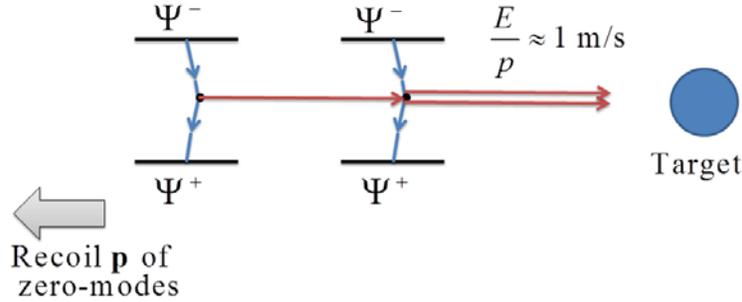

**Figure 3** – Gravitational vacuum fluctuations model of the emission of anomalous radiation by superconductors. In the "pumping" phase (not shown) the high-frequency interaction of a superconductor or a coherent field with the gravitational vacuum fluctuations excites some of them from their symmetrical virtual mass pair state $\Psi^+$ to the corresponding antisymmetric state $\Psi^-$. In the subsequent decay, strongly off-shell virtual gravitons are emitted, which cause further stimulated emission in the bulk and propagate to the target (only one stimulated emission is shown here). The virtual masses recoil; their momentum is passed either to the vacuum or to the material of the emitter.

The idea that the vacuum state in quantum mechanics has non-trivial properties and contains fluctuations is well established, but there are strong general limitations on vacuum processes. The reality of vacuum fluctuations is demonstrated by the Casimir effect in quantum electrodynamics, yet vacuum forces are usually very small, and the principles of thermodynamics limit the use of the Casimir effect for energy extraction from the vacuum [20]. The vacuum fluctuations that appear in our model, however, are of a novel kind, are peculiar of gravity and act on a far larger scale. This is why we think they can lead to macroscopic effect when coupled to macroscopic quantum objects like superconductors.

The features of the zero-modes have been derived in a purely theoretical way, but the phenomenology of Podkletnov experiment gives some clues in this direction. The anomalous force beam acts on the targets with a force proportional to their mass and appears to carry an energy and momentum with ratio $E/p \approx 1$ m/s, i.e. strongly off-shell, in the language of quantum field theory. Natural candidates as components of the beam are therefore virtual gravitons, and their source needs to be at the same time massive on a elementary-particles scale ($10^{-13}$ kg) and dipolar in a quantum sense, because classical mass dipoles do not exist. The zero-modes meet these requirements exactly at a length scale compatible with excitation by a superconductor wave function ($10^{-9}$ m).

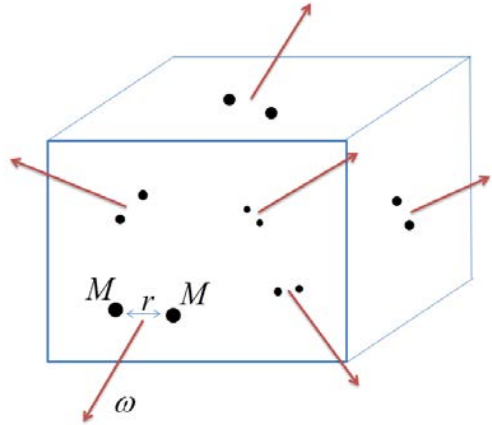

**Figure 4** – In any given volume of the emitter, the virtual mass couples emitting virtual gravitons (red) at a given frequency ω make up a continuum, according to the relation [7] $\omega = \frac{\Delta E}{\hbar} = \frac{2GM^2}{\hbar r}$ . Couples with larger mass have a larger size, and vice versa. For ω of the order of $10^7$ Hz (pumping frequency) and $r$ of the order of $10^{-9}$ m (coherent length of the superconductor), $M$ is of the order of $10^{-13}$ kg. This is large on the atomic scale; one of the consequences is that the recoil energy of the virtual masses is very small.

Recently there has been in astrophysics and cosmology much interest for dark matter and dark energy [14], but their density would be low at the local scale and it seems that at the local level an extraction of energy and momentum from this background would be hard – though there are proposals in this direction.

## 3. PHENOMENOLOGY OF MOMENTUM EXCHANGE AND EMITTER RECOIL IN THE VACUUM FLUCTUATIONS MODEL

The details of the interactions which occur in the emitter and lead to its recoil are quite complex. Further experimental trials will be necessary in order to obtain a clear picture of the phenomenology and devise possible ways to optimize and improve the effects.

The established theoretical relevant factors (within our model) are the following:

1. After the emission of the virtual gravitons the zero-modes recoil, with opposite momentum, but in the same direction of the emission, because of their negative mass.
2. The virtual gravitons are easily absorbed by light mobile targets, but tend to penetrate heavy targets or screens (this is confirmed experimentally; see Fig. 2).

3. The zero-modes, on the contrary, have a very small cross-section for scattering by ordinary matter. The scattering cross-section is possibly larger for coherent matter, but still unknown. Compare Sect. 4.1.
4. The zero-modes certainly cannot transfer their excess momentum to other zero-modes, because of their Lorentz invariance. In the rest system of a moving zero-mode, the vacuum will appear the same as if the zero-mode was not moving. (This resembles a continuous version of an Umklapp process in a crystal.)
5. The emission of virtual gravitons is always a process of interaction with the targets. The virtual graviton beam cannot propagate to infinity. The emission depends on the available targets. In this sense, there is no causal temporal relationship between emission and absorption, like for a real beam.

That said, the details of the process are influenced by several additional variables. A systematic treatment would be very long, because of the many possible alternatives and related uncertainties. We will only make some examples of how the various factors determine the emission, the recoil and therefore the possible applications to propulsion.

### 3.1 First major variable: spontaneous vs. stimulated emission

According to our model, the spontaneous emission of virtual gravitons in the decay of zero-modes does not have any preferred direction and cannot be influenced by electric fields. (In [6] we briefly mentioned that the electric field may contribute through the local-$\Lambda$ term to the pumping of the excited level. Here we give a quantitative estimate of this effect (Sect. 4.2). It turns out, however, that the field strength needed is very large.) Therefore in the absence of stimulated emission, or when stimulated emission is weak, the emission is isotropic, as in Fig. 5.1.

(An exception is a possible influence of the available targets, since the emission must be followed by absorption in a target; this is one of the conditions which can be relevant for applications to propulsion, also depending on whether the target is attached to the emitter or not, compare Sect. 3.2.)

If there are multiple stimulated emissions, one can easily check that any direction which is initially favored for some reason ends up being strongly dominant, due to the amplification effect (see for instance [21], p. 50). In the situation depicted in Fig. 5.3, for instance, the emission happens to be favored in the direction orthogonal to the emitter surface. This appears to be the case for Podkletnov's emitters, probably due to their ordered crystal structure.

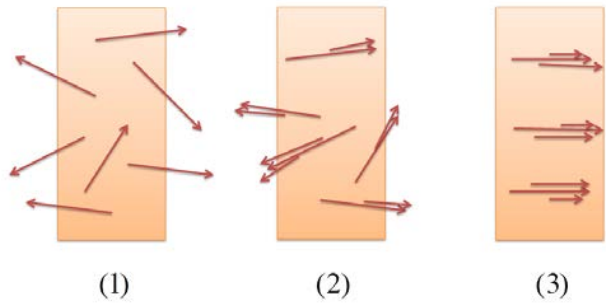

**Figure 5** – Spontaneous vs. stimulated emission of virtual gravitons in the superconducting emitter. (1) Only isotropic spontaneous emission. (2) Weak isotropic stimulated emission. (3) Strong stimulated emission, leading to cascades which greatly amplify a preferred emission direction (in this example, the direction orthogonal to the emitter surface).

## 3.2 Second major variable: prevailing emission towards the bulk of the emitter, or towards the outside

In order to illustrate this point, we do not start from any of the situations of Fig. 5, but from a simpler situation (Fig. 6) where the emission can be either one-directional or bi-directional, and originate either from all the bulk or from superficial layers. This may in turn happen because the emitter is not homogeneous, but has layers with different superconducting properties; the pumping process and the population inversion are mainly determined by the Cooper pairs density and its gradient [7]. Or the inhomogeneity could be due to the presence of stimulated emission; for instance, in Fig. 5.3 most emissions occur near one of the surfaces (but in Fig. 6 we do not take into account stimulated emission, for simplicity).

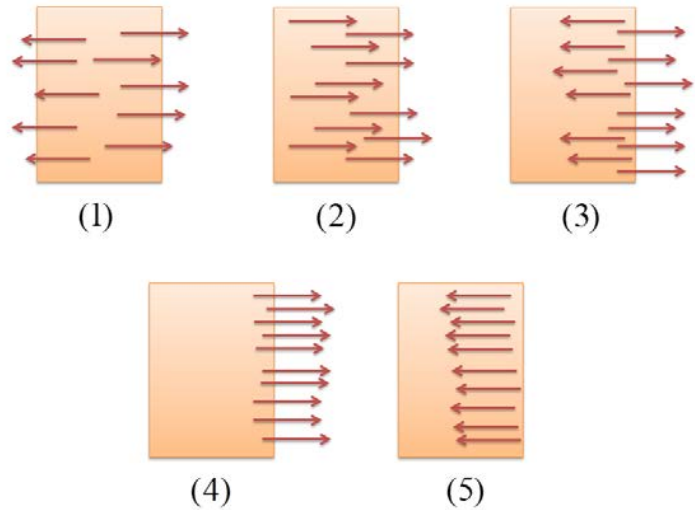

**Figure 6** – Possible combinations of one-directional or bi-directional, homogeneous or inhomogeneous emission (apart from trivial left-right exchanges). (1) Bi-directional and homogeneous. (2) One-directional and homogeneous. (3) Bi-directional and inhomogeneous. (4) One-directional and inhomogeneous, towards the outside. (5) One-directional and inhomogeneous, towards the bulk of the emitter.

Now in all these situations we have to consider the different effect of the recoil of the zero-modes. The simplest case is that of Fig. 6.4: if the emission layer is thin and close to the surface, neither the emitted virtual gravitons nor the recoiling zero-modes can release their momentum to the emitter (Fig. 7.a). This may be the case of Podkletnov, if the absence of recoil is confirmed.

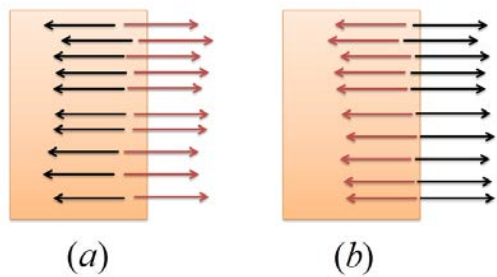

**Figure 7** – Recoiling zero-modes in the cases 4 and 5 of the previous picture. The virtual gravitons (red arrows) are easily absorbed by normal matter and release their momentum; the zero-modes can be scattered by coherent matter, but the scattering cross-section is still unknown. The momentum carried by the zero-modes (black arrows) is opposite to their velocity. In **(a)** there is no way for the virtual gravitons or for the zero-modes to pass momentum to the emitter, therefore no recoil. In **(b)** the emitter "shoots virtual gravitons on itself" and may in principle also receive some momentum from the recoiling zero-modes. Remember, however, that the recoil velocity is very small (Sect. 4.1), so in practice it is irrelevant as soon as the emitter is in motion.

In the case of Fig. 7.b the emitter is "shooting virtual gravitons on itself"; if the interaction with the recoiling zero-modes is small, then the total effect is recoil to the left. The same would happen for (2) and (3) in Fig. 6, with recoil respectively to the right and to the left. One of these cases 5.2, 5.3, 7.b could correspond to the effect observed by Poher.

Since these are only a few special examples, it is clear that in general the phenomenology can be quite complex. Finally, we observe that if the recoil of the emitter is caused by virtual gravitons generated and absorbed in the emitter itself (emitter "shooting on itself"), then this recoil does not need an external target, and the momentum is completely balanced by the recoiling zero-modes. On the contrary, if the virtual gravitons are generated in the emitter but absorbed by an external target, and the recoiling zero-modes do not interact with the emitter, then there will be no recoil of the emitter. All this is clearly crucial for applications to propulsion. Also remember that the experiments have been made so far in the laboratory, where the virtual gravitons beam can always "dump" its momentum on an external target.

## 4. DETAILS AND IMPROVEMENTS OF THE THEORETICAL MODEL

Formal work on this matter is in progress. We only mention here the main ideas.

### 4.1 Scattering cross sections of zero-modes

For an order of magnitude estimate of the gravitational scattering cross section of a zero-mode on an atom, we can consider the zero-mode as a negative mass of the order of $10^{-13}$ kg and the atom as a positive mass of the order of $10^{-27}$ kg. In the laboratory system, a zero-mode initially at rest has a very small recoil velocity ($10^{-14}$ m/s), therefore the scattering can only occur when the emitter is already in motion. Since, however, the zero-modes are Lorentz-invariant, they have a continuum distribution of initial velocities, and a complete vectorial description of the scattering is very complex. For small velocities, the cross section is of the order of $10^{-18}$ m$^2$. For large velocities, the excitation probability of the zero-modes by interaction with the superconductor is affected by the transit time and the relativistic frequency shift.

The scattering cross section for collisions between zero-modes and coherent matter might be larger, as mentioned, in analogy with the pumping effect of the superconductor. Generally speaking, the role of coherence in a scattering with $N$ particles is that of increasing the cross sections by a factor $N$ due to the coherent sum of amplitudes. The phenomena of coherent pumping and coherent scattering are quite different, however. With respect to pumping, a zero-mode behaves like a two-state system; its inertia and the inertia of coherent matter do not seem to be relevant in this case. On the other hand, inertia is relevant for scattering, and both types of coherent matter we considered (superconducting electron pairs and electromagnetic field, compare also Sect. 4.2) have very little inertia.

### 4.2 Electric and magnetic field strength required for a pumping effect comparable to that of superconductors

The local $\Lambda$-term, or vacuum energy term, which is able to excite the zero-modes, receives from superconductors with large pair density gradients contributions of the order of $10^6 - 10^8$ J/m$^3$. The electric and magnetic energy densities are given respectively by $\varepsilon_0 \mathbf{E}^2/2$ and $\mathbf{B}^2/(2\mu_0)$. If the fields are in the low-frequency limit, in states where the photon number uncertainty is much larger than the phase uncertainty, then they contribute to the local $\Lambda$-term. The strengths required to obtain a density of $10^6 - 10^8$ J/m$^3$ are, however, very large: $10^9$ V/m and 10 T, respectively. This explains why the anomalous emission has been only observed, until now, with superconductors. Note that the $\Lambda$-term must also oscillate, with a frequency of at least $10^6$ Hz, in order to efficiently excite the zero-modes. This appears to exclude any role of the pure $\mathbf{B}$ field. An induction $\mathbf{E}$ field may instead play a significant role. It is straightforward to compute, in dependence on the current, the geometric requirements for high-frequency coils which are able to give a field of $10^9$ V/m at $10^6$ Hz. The technical viability is a more subtle engineering matter.

### 4.3 Analogies with other quantum phenomena

In quantum field theory there are several well-known examples of phenomena described as exchange and propagation of gravitons. The corresponding amplitudes can be computed in a standard way at the lowest perturbation order, where the problem of non-renormalizability is not present. (In the effective-QFT approach [22], the non-renormalizability issue is removed also at higher orders.) Some of these phenomena are depicted in Table 1. The table also lists similarities and differences with respect to the graviton exchange hypothesized in our model. The first diagram in Table 1 represents a scattering of two massive particles with exchange of one virtual graviton. The second diagram contains two integrations over time and allows to compute the static interaction potential of two massive particles, as proven through a formula [23] which has also been used for higher-order computations [24] and for computations on a lattice [25]. The third diagram simply shows a spontaneous graviton emission and serves mainly as a comparison with the other two diagrams of the table and with those of Fig. 8.

| QUANTUM PHENOMENON | SIMILARITIES | DIFFERENCES |
|---|---|---|
| 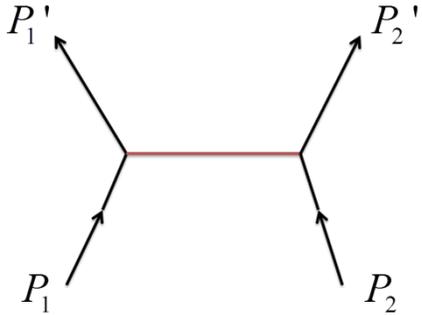<br>**Scattering** | • Exchange of one virtual particle with large momentum and small wavelength. | • Short range.<br>• Both masses are real and positive. |
| 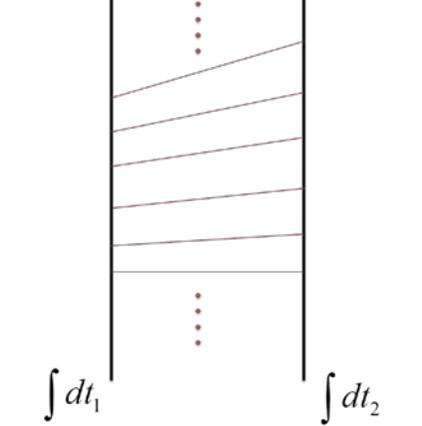<br>**Static interaction** | • Long range.<br>• Exchange of virtual particles with $E/p \approx 1$ m/s. | • Virtual particles have large wavelength.<br>• Both masses are real and positive. |
| 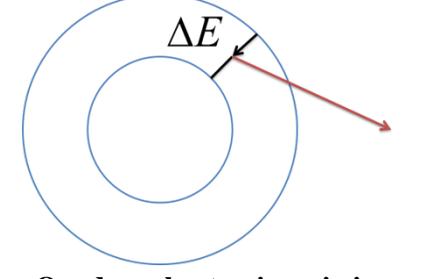<br>**Quadrupole atomic emission** | • Energy of emitted particle is equal to difference between energy levels. | • Emitted graviton is real and has spin 2. |

**Table 1** – Similarities and differences between known quantum phenomena and the interactions hypothesized by our model. See explanations in the main text. In the diagrams the red lines represent massless gravitons and the black lines massive particles. The possible occurrence of stimulated emission has been disregarded in these comparisons. Note that for tree Feynman diagrams involving only internal graviton lines, there is a complete analogy between electromagnetism and gravity, through the Einstein-Maxwell equations in "Adler gauge" ([7], Appendix 2).

These analogies can guide us in improving the theoretical model sketched in Fig. 3. The computations, reported in Ref. [7], of the probabilities of the main phases of the process (virtual graviton emission, propagation and absorption) only give a first approximation. For instance, in [7] we evaluated the probability of spontaneous emission just by substituting the virtual wavelength in the usual expression of the Einstein *A*-coefficient; this should be replaced by a full computation of the amplitude of the "distant scattering" process whose electromagnetic analog is represented in Fig. 8.b, also including an integration over the many possible virtual-mass initial states (Fig. 4).

For a more detailed comparison, consider an atom in an excited state with angular momentum *l*=1 (in units $h/2\pi$), which decays to its ground state emitting a real photon. If the excitation energy is $\Delta E$ and the four-momentum of the photon is $P=(E,\mathbf{p})$, then we have $E^2-p^2c^2=0$. The recoil momentum and recoil energy are very small, therefore $E \approx \Delta E$. Netx consider an atom in an excited state with *l*=0. A dipole transition with emission of a real photon is forbidden, but a de-excitation following a collision with, for instance, a proton, is possible (Fig. 8.a). In the collision a virtual off-shell photon is exchanged, which carries the appropriate four-momentum $P=P_2'-P_2$ and zero angular momentum.

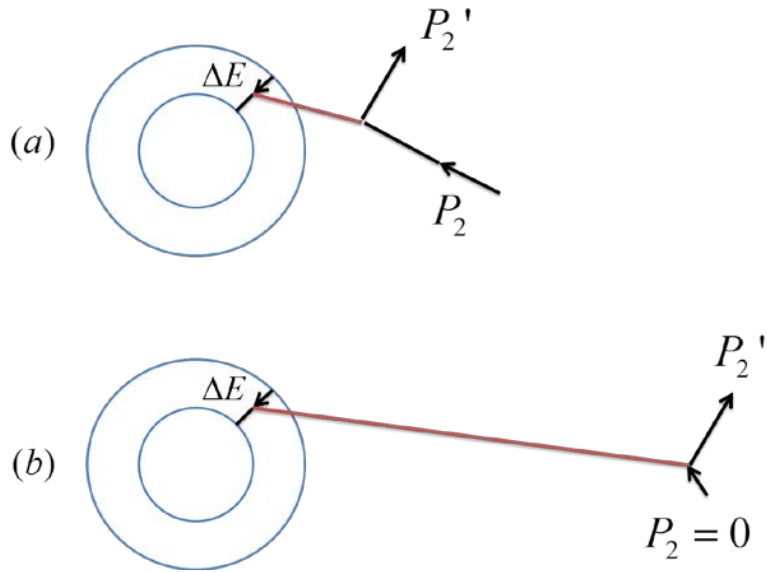

**Figure 8 – (a)** De-excitation of an atom in a state with *l*=0 after collision with an ion. An off-shell photon is exchanged, which carries part of the de-excitation energy to the ion. **(b)** Same as in (a), but at larger distance and with an ion which is initially at rest ("far virtual collision"). In both cases, the diagram is incomplete, because the ion also exchanges energy and momentum with the atomic nucleus; this makes the far virtual collision much less probable. If, however, the diagrams are thought to represent the de-excitation of a couple of gravitational zero-modes, then they are complete and the long-range interaction is not much less probable than the short-range interaction.

Finally, consider the process of Fig. 8.b, which we might call "de-excitation by far virtual collision". The balance of four-momentum and angular momentum is similar to that of diagram 8.a, except for the fact that the colliding particle is initially at rest. If we suppose that this particle is a proton, then we can use non-relativistic expressions for $E_2$', $p_2$' and we find $E_2'/p_2'=v/2$, where $m$ is the proton mass and $v$ its velocity after the collision. In fact, however, there is no collision, since the proton never approaches the atom. The process is kinematically allowed, but must be strongly suppressed at distances much larger than the atomic size; otherwise, as soon as there are available protons somewhere, the excited state would decay quickly even though the dipolar transition is forbidden.

Clearly, the difference between the two processes of Fig. 8 must be in the propagation amplitude of the virtual photon, which decreases fast at large distances. This is due, in turn, to the short range of the force: although it is not apparent from our diagrams, if we are handling with an atom the interaction will involve both the electron and the nucleus, resulting of the Van der Waals type. If, on the contrary, the diagrams represent a gravitational interaction, we may expect that the propagation amplitude of the virtual graviton decreases with the distance, but only according to a power law. This is one of the reasons why we expect that a "far collision" may actually happen in the decay of the zero-modes. The other reason is that the zero-modes, unlike the excited states of an atom, form a continuum, and this increases the total probability of a process involving them, because starting from the same pumping quantum $hf$ there are many possible intermediate states which lead to the same final process.

## 5. CONCLUDING REMARKS

We are aware that the concepts outlined in this paper are groundbreaking and still hard to understand. We believe, however, that they are scientifically sound and based on experimental results and known theory. Most of them are rigorously proven, but not all are settled within standard theoretical methods. In particular, weak points of the model are the definition of the structure of the gravitational vacuum including the zero-modes and the non-perturbative computation of the pumping effect of the Λ–term; these are very complex issues, whose solution will probably take some time and the combined efforts at many a theoretician. We regard gravity-superconductors interactions as a crucial test for a quantum theory of gravitation, and actually as a setting where the theory can mature (as the rest of quantum physics did) in strict connection with the experiments.

Finally, we would like to append an update of ongoing discussion on two recent issues concerning gravity-superconductors interactions.

**1. Possibility that the superconducting emitters go normal during the current pulse.** Some observers have pointed out that there is no proof that the material of the emitters remain superconducting during the discharge. The standard techniques which allow such a check are very hard to implement in the experimental conditions of Podkletnov and Poher. If we cannot be sure that the emitters are superconducting, the objection goes, than the whole interpretation of the phenomenon is questionable. Poher has recently replied that he is aware of the problem, and that also for this reason an independent replication of his experiments failed in 2012. This replication attempt used too large densely sintered pellets, a much too large current, and an inadequate cooling down method. A second replication by the same team in 2013, which took into account his suggestions, was successful. According to Poher, for these same reasons he stopped using Types I, II, and III emitters made of small compact sintered modules after 2007. They are limited in maximum discharge current and limited in performance because they are destroyed by the propelling force (cracks), and they are quite difficult to cool down correctly. The new emitters (Type V to X) are highly porous, and their thin (microns) grains have a high "skin surface-to-mass" ratio, so they are almost completely surrounded by liquid nitrogen. Concerning Podkletnov, he reported effects in the temperature range 50-70 K (liquid helium cooling). This should be cold enough to prevent the emitter from going over $T_c$ (92 K) in the discharge. Also consider that the duration of his discharges was only 0.1 – 1 microseconds. The (measured) critical current of the melt-textured material of the emitter was large, about 50000 $A/cm^2$, and the emitter surface was ca. 75 $cm^2$.

**2. Revision of analysis of the beam propagation velocity.** The data and analysis reported in [17] also roused considerable feedback. An elaboration of this feedback is in progress. Although the measurements appear to be robust, the phenomenon is startling, complex and difficult to understand, like other effects of this kind. The theoretical analysis of [17] should almost certainly be improved as follows: since the piezoelectric sensor is actually operating in resonating mode, its response should be re-computed and the exchange of energy and momentum with the beam does not require local conservation of energy, because of the external power supply at resonance.